\documentclass[prl,two column]{revtex4}
\usepackage{amssymb}

\usepackage{amsmath}

\input{tcilatex}

\begin{document}

\title[Infinite order excitonic Bloch equations]{Infinite order excitonic
Bloch equations for asymmetric nanostructures}
\author{M. Hawton}
\affiliation{Department of Physics, Lakehead University, Thunder Bay, ON, Canada, P7B 5E1}
\author{M. M. Dignam}
\affiliation{Department of Physics, Queen's University, Kingston, ON, Canada, K7L 3N6}

\begin{abstract}
We present a new exciton-based formalism for calculating the coherent
response of asymmetric semiconductor multiple quantum well structures to
ultra-short optical pulses valid to infinite order in the optical field and
including the self-generated intraband fields. We use these equations to
calculate and explain the oscillations with time delay of peaks in the
spectrally-resolved degenerate four wave mixing signals from biased
semiconductor superlattices, obtaining good agreement with experiment.
\end{abstract}

\maketitle

Over the last decade, there has been a great deal of interest in the
nonlinear optical response of semiconductor nanostructures to ultra-short
optical pulses. \ Many phenomena have been treated using the dynamics
controlled truncation (DCT) theory of Axt and Stahl \cite{AxtStahl94},
whereby the response is expanded in powers of the exciting laser field, and
is formulated to include n-point correlations \cite{AxtStahl94,
HawtonNelson98,DignamHawton03}. \ This theory has been very successful, and
correctly deals with the intra- and inter- excitonic correlations to
arbitrary order in the optical field. \ However, it is intrinsically a
perturbative technique, while the theoretical description of many
interesting nonlinear processes requires a non-perturbative approach. \ 

The semiconductor Bloch equations (SBE's) in the Hartree-Fock (HF)
approximation provide a very compact description of dynamics to infinite
order in the optical field \cite{HaugKoch95}. \ They have been used to
describe Rabi splitting near zero detuning \cite{SchmittRinkChemlaHaug88}
and the ac Stark effect \cite{BinderKoch91}. \ It has been shown \cite%
{AxtBartelsStahl96,Bolivar97}, however, that the HF factorization of
four-particle correlation functions into products of two-particle interband
polarizations and electron and hole populations incorrectly predicts that
intraband correlations lose any excitonic signature as time progresses, and
act simply as free electron-hole pairs at long times.

In recent years, there has been a great deal of work on the nonlinear
optical properties of asymmetric coupled quantum well systems such as biased
semiconductor superlattices (BSSL's) \cite%
{AxtBartelsStahl96,Bolivar97,DignamHawton03,Lyssenko97,Shapiro,Liu,Shah,Dignam99}%
. \ Of particular interest has been the interplay of the interband and
intraband response in two-beam degenerate four-wave mixing (FWM)
experiments, where ultra-short optical pulses with wave vectors $\mathbf{k}%
_{1}$ and $\mathbf{k}_{2}$ separated by a delay time $\tau _{21}$ impinge on
a sample \cite{AxtBartelsStahl96,Bolivar97,DignamHawton03,Lyssenko97,Shapiro}%
. \ The interband polarization is responsible for pump-probe, four-wave
mixing and higher order signals, while the intraband polarization describes
the coherent motion of the electrons \emph{within} each band. \ In a BSSL,
the electron energy levels form the so-called Wannier-Stark ladder (WSL): $%
E_{n}=E_{0}+n\hslash \omega _{B}$, where $\omega _{B}\equiv eE_{DC}d/\hbar $
is the Bloch frequency, $d$ is the superlattice period, and $E_{DC}$ is the
applied along-axis DC electric field. \ Excitation of a coherent
superposition of WSL states results in oscillations in the intraband
polarization at $\omega _{B}$. \ This signature of Bloch oscillations (BO's)
has been seen, not only in the THz radiation generated by the intraband
polarization, but also in the time-resolved and time-integrated degenerate
four-wave mixing signals (TRFWM and TIFWM respectively). \ These features
can all be qualitatively understood using perturbative approaches such as
DCT \cite{AxtBartelsStahl96,DignamHawton03}. \ 

Recently, BO's have been observed in \emph{spectrally-resolved} four-wave
mixing (SRFWM) experiments \cite{Lyssenko97,Shapiro}, where the peak
energies of excitonic resonances were found to oscillate with the time
delay, $\tau _{21}$, at the Bloch frequency. \ Although it has been
suggested that these oscillations result from the influence of the excitonic
intraband dipole on the excitonic energies \cite{Lyssenko97,Shapiro,Dignam99}%
, the precise mechanism has never been clear and there has been some
controversy regarding this interpretation \cite{Liu}. \ It is known that
third-order DCT-type calculations do not yield significant peak oscillations %
\cite{DignamHawton03}. \ In fact, as we shall show, peak oscillations
similar to the experimentally-observed ones will only arise from a formalism
that is infinite order in the optical field.

In this letter, \emph{we present a new nonlinear response formalism, valid
to infinite order in the optical field, that combines the essential features
of both the SBE's and DCT theory} for asymmetric systems. \ Our approach
relies on the fact that, in such systems, exciton-exciton interactions are
dominated by a dipole-dipole interaction \cite{DignamHawton03} and phase
space filling (PSF) effects are relatively insignificant for the range of
carrier densities considered \cite{PSFcomment}.\ \ In what follows, we
present our theory and calculate the TIFWM and SRFWM signals for a BSSL. We
obtain oscillations in the SRFWM peak energies that are in good agreement
with experimental results \cite{Lyssenko97,Shapiro}.\ \ Furthermore, we find
that the system is so nonlinear, that SRFWM and TIFWM results cannot be
adequately described using a third order DCT approach, \emph{even at
moderate densities} (\symbol{126}$3\times 10^{9}/$cm$^{2}$).

We formulate our theory in the basis of the \emph{excitonic states} of the
BSSL \emph{in the presence of the applied dc field}, $E_{DC}$. \ While
recent work shows that treatment of PSF in an exciton basis is difficult and
subtle \cite{DignamHawton03,BosonRefs}, these problems do not arise for the
densities considered here \cite{PSFcomment}. \ The excitons are
characterized by the quantum numbers $(\mu ,n)$, where $\mu $ describes the
internal motion and $\mathbf{K}_{n}$ is the center of mass wave vector. \
This wave vector determines the direction of the optical radiation emitted
by the exciton. \ In degenerate FWM experiments, incident optical pulses
create excitons with wave vectors $\mathbf{K}_{1}\equiv \mathbf{k}_{1}$and $%
\mathbf{K}_{-1}\equiv \mathbf{k}_{2}$. \ These excitons produce an \emph{%
intraband} polarization grating with wave vectors $\mathbf{K}_{-2}=\pm
\left( \mathbf{k}_{1}-\mathbf{k}_{2}\right) $ that scatter the
optically-created excitons into wave vectors $\mathbf{K}_{3}\equiv 2\mathbf{k%
}_{1}-\mathbf{k}_{2}$ and $\mathbf{K}_{-3}\equiv 2\mathbf{k}_{2}-\mathbf{k}%
_{1}$. \ Considering all scattered excitons, the intraband grating is
described by wave vectors\ $\mathbf{K}_{m}=\frac{m}{2}\left( \mathbf{k}_{1}-%
\mathbf{k}_{2}\right) $ ($m$ even) and excitons are scattered into all wave
vectors $\mathbf{K}_{n}=\frac{1}{2}\left[ \left( n+1\right) \mathbf{k}%
_{1}-\left( n-1\right) \mathbf{k}_{2}\right] $ ($n$ odd) \cite%
{DignamHawton03}. \ Henceforth, we refer to signals in the FWM direction, $%
\mathbf{K}_{-3},$ as FWM signals, although they are to infinite order in the
optical field.

Thus, for our BSSL, the Hamiltonian is given by \cite{DignamHawton03}

\begin{eqnarray}
H &=&\sum_{\mu ;\,n=-n_{0}}^{n_{0}\,by\,2}\hbar \omega _{\mu }B_{\mu
,n}^{\dagger }B_{\mu ,n}-V\sum_{n=\pm 1}\mathbf{E}_{n}^{opt\ast }\cdot 
\mathbf{P}_{n}^{inter}  \notag \\
&&+V\sum_{m=-2n_{0}}^{2n_{0}\;by\;2}\left[ \frac{1}{2\varepsilon }\mathbf{P}%
_{-m}^{intra}-\mathbf{E}_{ext}^{THz}\delta _{m,0}\right] \cdot \mathbf{P}%
_{m}^{intra}\text{,}
\end{eqnarray}%
where $V$ is the system volume, $\varepsilon $ is the permittivity of the
BSSL, $\hbar \omega _{\mu }$ is the energy and $B_{\mu ,n}^{\dagger }$ is
the creation operator of the exciton (in the DC field) in the state $(\mu
,n) $ (for $n$ odd). \ The incident optical field is given by $\func{Re}%
\left[ E_{n}^{opt}\left( t\right) \right] =\func{Re}\left[ {\LARGE %
\varepsilon }_{n}^{opt}\left( t\right) e^{-i\omega _{c}t}\right] $, where $%
\omega _{c}$ is the laser central frequency, ${\LARGE \varepsilon }%
_{n}^{opt}\left( t\right) $ is the complex temporal envelope, and $\mathbf{E}%
_{ext}^{THz}\left( t\right) $ is an external THz field. \ The interband
polarization with wave vector $\mathbf{K}_{n}$ ($n$ odd) is $\mathbf{P}%
_{n}^{inter}\mathbf{=}\frac{1}{V}\sum_{\mu }\mathbf{M}_{\mu }B_{\mu
,-n}^{\dagger }+h.c.$, where $\mathbf{M}_{\mu }$ is the excitonic interband
dipole matrix element. \ Finally, the intraband polarization with wave
vector $\mathbf{K}_{m}$ ($m>0$ and even) is $\mathbf{P}_{m}^{intra}=\frac{1}{%
V}\sum_{\mu ,\mu ^{\prime }}\sum_{n}\mathbf{G}_{\mu ,\mu ^{\prime }}B_{\mu
,n}^{\dagger }B_{\mu ^{\prime },m+n}$, where $\mathbf{P}_{-m}^{intra}=%
\mathbf{P}_{m}^{intra\ast }$ and the sum over $n$ is from $-n_{o}$ to $%
n_{0}-m\,$by 2, and $n_{o}$ is an odd, positive integer \cite{DignamHawton03}%
. \ The index, $n_{o}$ serves as a truncation point for the sum; if $n_{o}$
is infinite then the expression is exact. \ In actual calculations we
increase $n_{o}$ until convergence is obtained. \ 

We use the Heisenberg equations to determine operator dynamics and truncate
the hierarchy of equations by factoring the three-exciton (six-particle)
correlation functions into one- and two-exciton correlation functions. \
This factorization is very different than that used in the SBEs, as it
retains the intra-excitonic electron-hole correlations and the long range
exciton-exciton correlations. \ It has been shown to be very accurate for
the calculation of degenerate FWM signals to third order in BSSL's \cite%
{DignamHawton03}. \ We account for dephasing and decoherence
phenomenologically via the interband and intraband dephasing times, $T_{\mu
} $ and $T_{\mu \nu }$ respectively. The dynamical equation for the $%
\left\langle B_{\mu ,n}^{\dagger }\right\rangle $ thus becomes:%
\begin{eqnarray}
i\hbar \frac{d\left\langle B_{\mu ,n}^{\dagger }\right\rangle }{dt}
&=&-\hbar \left( \omega _{\mu }+\frac{i}{T_{\mu }}\right) \left\langle
B_{\mu ,n}^{\dagger }\right\rangle +\mathbf{E}_{n}^{opt\ast }\cdot \mathbf{M}%
_{\mu }^{\ast }  \notag \\
&&+\sum_{\nu ;\,\,k=-n_{0}}^{n_{0}\;by\;2}\mathbf{E}_{n-k}^{intra\ast }\cdot 
\mathbf{G}_{\nu ,\mu }\left\langle B_{\nu ,k}^{\dagger }\right\rangle ,
\label{<B+>}
\end{eqnarray}%
where $\mathbf{E}_{-m}^{intra}\equiv -\frac{1}{\varepsilon }\left\langle 
\mathbf{P}_{m}^{intra}\right\rangle +\mathbf{E}_{ext}^{THz}\delta _{m,0}$ is
the total intraband field with wave vector $\mathbf{K}_{m}$, and an equation
similar to Eq. (\ref{<B+>}) is used to obtain the $\left\langle B_{\mu
,n}^{\dagger }B_{\nu ,j}\right\rangle $ needed in the calculation of $%
\left\langle \mathbf{P}_{m}^{intra}\right\rangle $.

Note that in the above equations, the $\mathbf{K}=0$ \emph{internal}
intraband field interacts with the excitons in exactly the same way as an 
\emph{external} THz field. \ The dc component of this field renormalizes the
excitonic energy, $\hbar \omega _{\mu }$, while its THz component
dynamically couples levels with different $\mu $ \cite{Dignam99}. \ As
discussed below, it is largely the $\mathbf{K}=0$ internal field that yields
the peak oscillations in the SRFWM signal. \emph{\ }A DCT calculation to any
order will not yield these oscillations because the equation for $%
\left\langle B_{\mu ,-3}^{\dagger }\right\rangle ^{(n)}$ ($\left\langle
B_{\mu ,-3}^{\dagger }\right\rangle $ to order $n$) does not contain a term
proportional to $\mathbf{E}_{0}^{intra}\left\langle B_{\mu ,-3}^{\dagger
}\right\rangle ^{(n)}$.

We now present calculated results for the GaAs/Ga$_{0.7}$Al$_{0.3}$As
superlattice used in recent experiments \cite{Lyssenko97,Shapiro}: the well
widths and barrier widths are $67$\AA\ and $17$\AA\ respectively, the dc
electric field is 15 kV/cm, and there is no\emph{\ external }THz field. \
The single-particle BO period of this system is $\tau _{B}=328$ fs, with a
corresponding WSL energy spacing of $\hbar \omega _{B}=12.6$ meV. \ The
system is excited by a Gaussian optical pulse with a temporal FWHM of 90 fs.
\ The dephasing times are taken to be $T_{inter}=T_{\mu }=1.0\,\ $ps and $%
T_{intra}=T_{\mu \nu }=1.5\,\ $ps, while the excitonic population decay
time, $T_{\mu \mu }$, is taken to be infinite \cite{Tcomment}.\ \ As in the
experiments \cite{Lyssenko97,Shapiro}, the exciton areal density ranges from 
$10^{9}$ to $10^{10}$/cm$^{2}$. \ Only $1s$ heavy-hole excitons are
included, so that the single internal quantum number $\mu $ describes an
exciton in which the average along-axis electron-hole separation is
approximately $\mu d$ \cite{DignamHawton03}. \ Thus, for the exciton, $\mu $
plays the role that $n$ does for the single-particle WSL states except that
the excitonic WSL is not equally spaced \cite{DignamHawton03}. \ These
states are calculated using the method described in Ref. \cite{Shah}. \ The
neglect of excited in-plane excitonic states has been shown to be justified
for central laser frequencies below the $\mu =0$ WSL frequency \cite{Shah}.

Fig. 1 shows the TIFWM intensity as a function of delay time. \ The
oscillations are due to BO's of the excitonic wavepackets \cite%
{DignamHawton03,Dignam99}. \ The frequency of the oscillations is not
precisely given by $\omega _{B}$ for two reasons. \ First, the dc component
of the $\mathbf{K=}0$ internal THz field renormalizes the excitonic energies
by giving an effective dc field of $\mathbf{E}_{DC}+\mathbf{E}_{0}^{intra,DC}
$; because $\mathbf{E}_{0}^{intra,DC}$ is antiparallel to $\mathbf{E}_{DC}$
for $\omega _{c}<\omega _{0}$, the BO frequency is reduced, as observed. \
Second, excitonic effects alter the BO frequency by changing the WSL energy
spacings as mentioned above.

The large degree of nonlinearity in this system is seen in the TIFWM results
obtained using different truncation indices, $n_{o}$ (dotted lines in Fig.
1). \ For $n_{0}=3$, the TIFWM signal ($\mathbf{K}_{-3}$ direction) arises
entirely from $\mathbf{k}_{2}$ excitons that scattered off the $\mathbf{k}%
_{2}-\mathbf{k}_{1}$ grating. \ As $n_{o}$ is increased, the excitons can
scatter out of $\mathbf{K}_{-3}$ into wave vectors $\mathbf{K}_{\pm n}$ for
any odd $n\leq n_{0}$ and perhaps even scatter back into the FWM direction. $%
\ $Convergence is only obtained for $n_{o}\geq 9$ for this density, which
demonstrates that scattering into wave vectors as large as $\mathbf{K}_{\pm
9}$ is clearly important and thus a third-order DCT approach is not
adequate. \ The multiple scattering and the high degree of nonlinearity is
also evidenced in the fact that the intensities in the $\mathbf{K}_{-5}=3%
\mathbf{k}_{2}-2\mathbf{k}_{1}$ and $\mathbf{K}_{-7}=4\mathbf{k}_{2}-3%
\mathbf{k}_{1}$ directions, also shown in Fig. 1, are comparable to the FWM
intensity.

In Fig. 2 we plot the SRFWM signal for a sequence of delay times. \ The
spectral peaks are associated with different excitonic states (as indicated
in the figure). \ There are three key features of these spectra that we draw
attention to. \ First, the peaks do not occur precisely at the
single-excition energies, $\omega _{\mu }$; this is due to the
polarization-induced reduction of the applied dc field discussed above. \
Second, some of the peaks deviate very considerably from a Lorentzian shape.
\ Third, and of particular importance in this work, the peak positions
depend significantly on the time delay, $\tau _{21}$.

To see the peak oscillations more clearly, we plot in Fig. 3 the energy of
the $\mu =-1$ peak relative to the energy of the $\mu =-1$ excition as a
function of time delay for different exciton densities. \ As can be seen,
and in agreement with experiment, the peaks' energies oscillate as a
function of $\tau _{21}$ with period given approximately by $\tau _{B}$. \
To understand qualitatively these oscillations, we consider a simplified
model. \ We calculate the intraband fields to second order, use them in Eq. (%
\ref{<B+>}), and employ perturbation theory in all fields. \ The
oscillations can then be largely understood as arising from quantum
interference between multiple paths from the ground state to the excitonic
state $(\mu ,-3)$. \ In the inset to Fig. 2, we show the two lowest order
paths. \ In the left-hand path, an exciton with wave vector $\mathbf{K}_{-1}=%
\mathbf{k}_{2}$ and energy $\hbar \omega _{op}$ is created by the second
optical pulse; it then scatters off the $\mathbf{K}_{-2}$ grating into wave
vector $\mathbf{K}_{-3}$, absorbing a photon of energy $\hbar \omega
_{intra}^{G}\simeq 0,\pm \hbar \omega _{B}$. \ In the right hand path, the
exciton additionally interacts with the spatially uniform field, $%
E_{0}^{intra}\left( t\right) $, absorbing a photon of energy $\hbar \omega
_{intra}^{0}\simeq 0,\pm \hbar \omega _{B}$ before reaching the final state $%
\left( \mu ,-3\right) $. \ Including only the most important contributions
the left and right paths, their contributions to the SRFWM polarization near 
$\omega =\omega _{\mu }$ are approximately 
\begin{eqnarray}
P_{-3}^{\left( L\right) }\left( \omega \right) &\sim &\frac{1}{\left[ \omega
-\omega _{\mu }+i/T_{inter}\right] ^{2}}, \\
P_{-3}^{\left( R\right) }\left( \omega \right) &\sim &\frac{P_{-3}^{\left(
1\right) }\left( \omega \right) E_{0}^{intra}\left( \omega _{B}\right) }{%
\left[ \omega -\omega _{\mu }+i/T_{inter}\right] },
\end{eqnarray}%
where $E_{0}^{intra}\left( \omega _{B}\right) $ is the Fourier component of $%
E_{0}^{intra}\left( t\right) $ at the Bloch frequency. \ It is easily seen
that interference between these two contributions yields peaks in the SRFWM
spectra with energies that depend on the amplitude and phase of $%
E_{0}^{intra}\left( \omega _{B}\right) $. \ 

An approximate second order calculation for $E_{0}^{intra}\left( \omega
_{B}\right) $ yields 
\begin{equation}
E_{0}^{intra}\left( \omega _{B}\right) \sim e^{-i\omega _{B}\tau
_{21}/2}\cos \left( \omega _{B}\tau _{21}/2\right) .  \label{Eintra}
\end{equation}%
The oscillation of $E_{0}^{intra}\left( \omega _{B}\right) $ with time delay
is due to alternating constructive and destructive interference between the
Bloch oscillating intraband polarizations created by the first and second
pulses. \ Thus, the periodic oscillations in the SRFWM \emph{peak positions}
arise due to the quantum interference between two-photon and three-photon
processes. \ When the full equations are solved, one finds that higher-order
nonlinearities play an important role, yielding an oscillation-amplitude
that depends nonlinearly on exciton density. \ In the inset to Fig. 3, we
show the amplitude of the peak oscillations (defined to be the energy
difference between the first dip after $\tau _{21}=0$ and the peak that
follows) as a function of density for three different central laser
frequencies. \ Note that at low densities, the oscillation amplitude is
independent of density. \ In this regime, the peak oscillations are due to
small effects that occur even to third order in the optical field. \ At
higher densities the amplitudes increase superlinearly with density due to
multiple scatterings from the intraband polarization grating. \ At the
highest density of $9.3\times 10^{9}$ cm$^{-2}$, the sudden change in peak
energy near $\tau _{21}=0.16$ ps results from a splitting of the $\mu =-1$
peak, similar to that seen in Fig. 2 for the $\mu =0$ peak for $\tau
_{21}=0.34$ ps.

Most of the features of the calculated peak oscillations agree with
experiment: the phase, frequency, dependence on $\omega _{c}$, and large
amplitude at $\tau _{21}=0$, are all in general agreement \cite%
{Lyssenko97,Shapiro}. \ However, the amplitude of the peak oscillation is
considerably smaller than that obtained experimentally. \ For example, for $%
\omega _{c}=\omega _{0}-2.27\omega _{B}$, and a density of 10$^{10}$ cm$%
^{-2} $, the calculated amplitude is roughly 0.6 meV, while the experimental
amplitude is approximately 2.3 meV \cite{Shapiro}. \ We believe that this
difference is largely due to the effects of the screening of the
exciton-exciton interactions via incoherent carriers. \ As discussed in Ref. %
\cite{Shapiro}, this plasma screening will considerably decrease the
dielectric constant at the Bloch frequency. \ We find, for example, that a
decrease in $\varepsilon /\varepsilon _{o}$ from 12.5 to 8 can result in in
increase in the the peak-oscillation amplitude by a factor of 3 or more,
bringing the calculated and experimental results into quite good agreement.
\ For simplicity, we have used the static value of 12.5, but plan to use a
dynamic model in future calculations.

In summary, we have developed a new infinite order excitonic system of
equations for calculating the nonlinear optical response of asymmetric
semiconductor nanostructures. \ We have used these equations to reproduce
for the first time the experimentally-observed \cite{Lyssenko97,Shapiro}
oscillations in the SRFWM signals from BSSL's.

ACKNOWLEDGEMENTS

We wish to thank Karl Leo and Lijun Yang for valuable discussions. \ This
work was supported in part by the Natural Sciences and Engineering Research
Council of Canada.

\bigskip

FIGURE CAPTIONS

\begin{description}
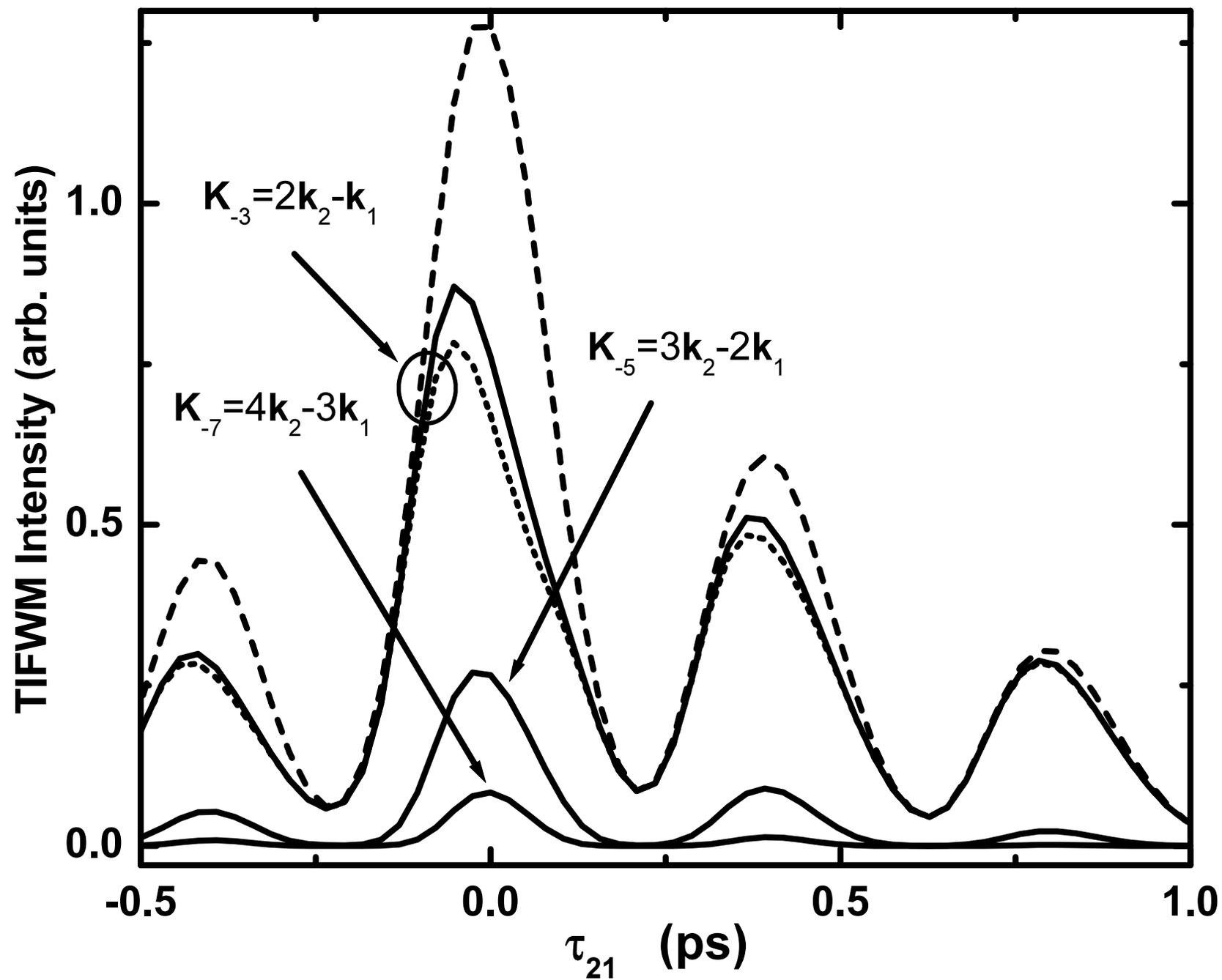
\item FIG. 1.\qquad The time-integrated intensity versus $\tau _{21}$ for $%
\omega _{c}=\omega _{0}-2.27\omega _{B}$, $n_{o}=13$, and a density of $%
6.36\times 10^{9}$ cm$^{2}$ for the three different directions, $\mathbf{K}%
_{n}$ indicated. \ For $\mathbf{K}_{-3}$, results for $n_{0}=3$ (dash) and $5
$ (dot) are also plotted.

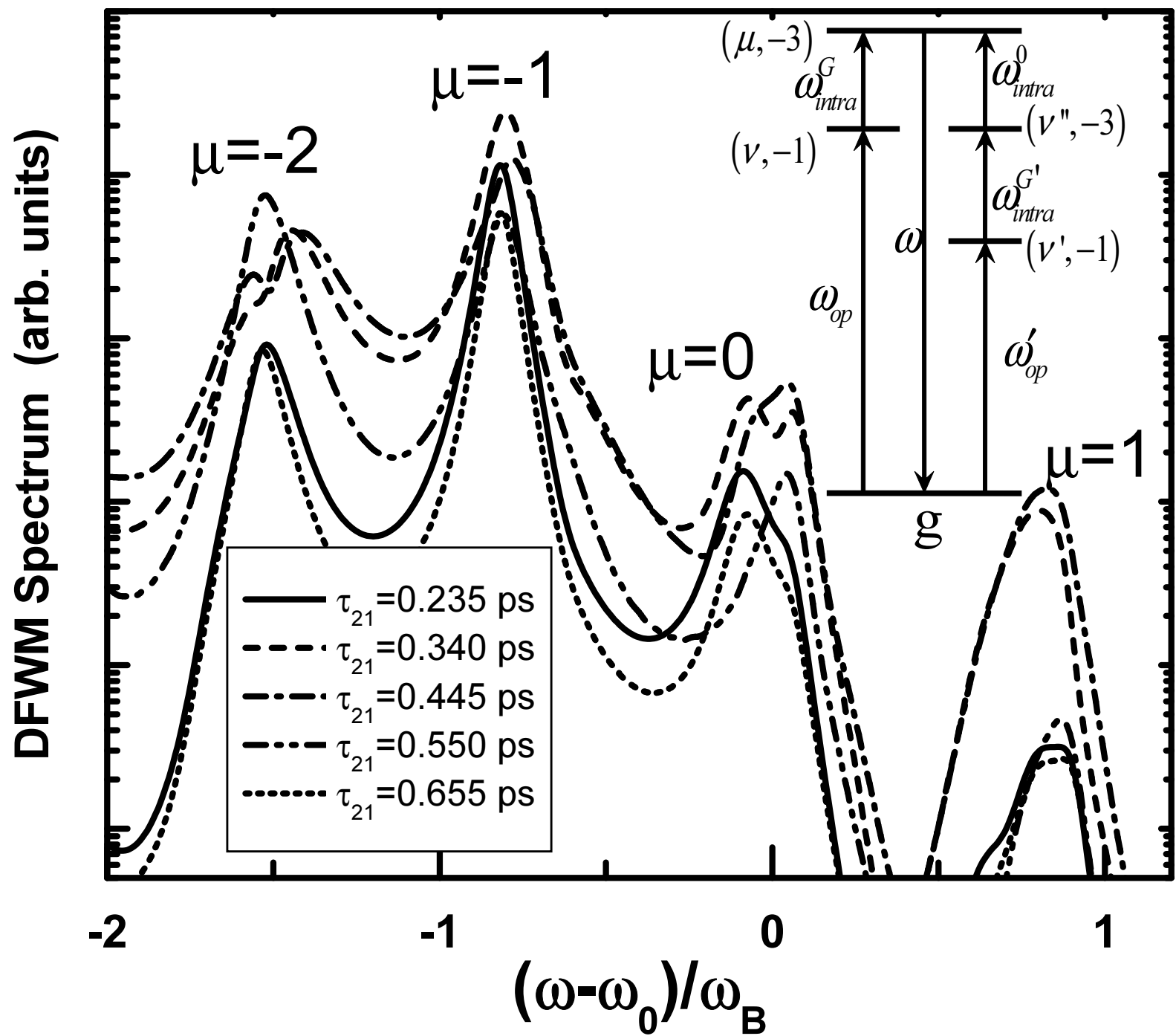
\item FIG. 2.\qquad SRFWM intensity versus frequency at $\omega _{c}=\omega
_{0}-2.27\omega _{B}$ and a density of $9.3\times 10^{9}$ cm$^{2}$ for a
sequence of delay times. \ The inset is described in the text.

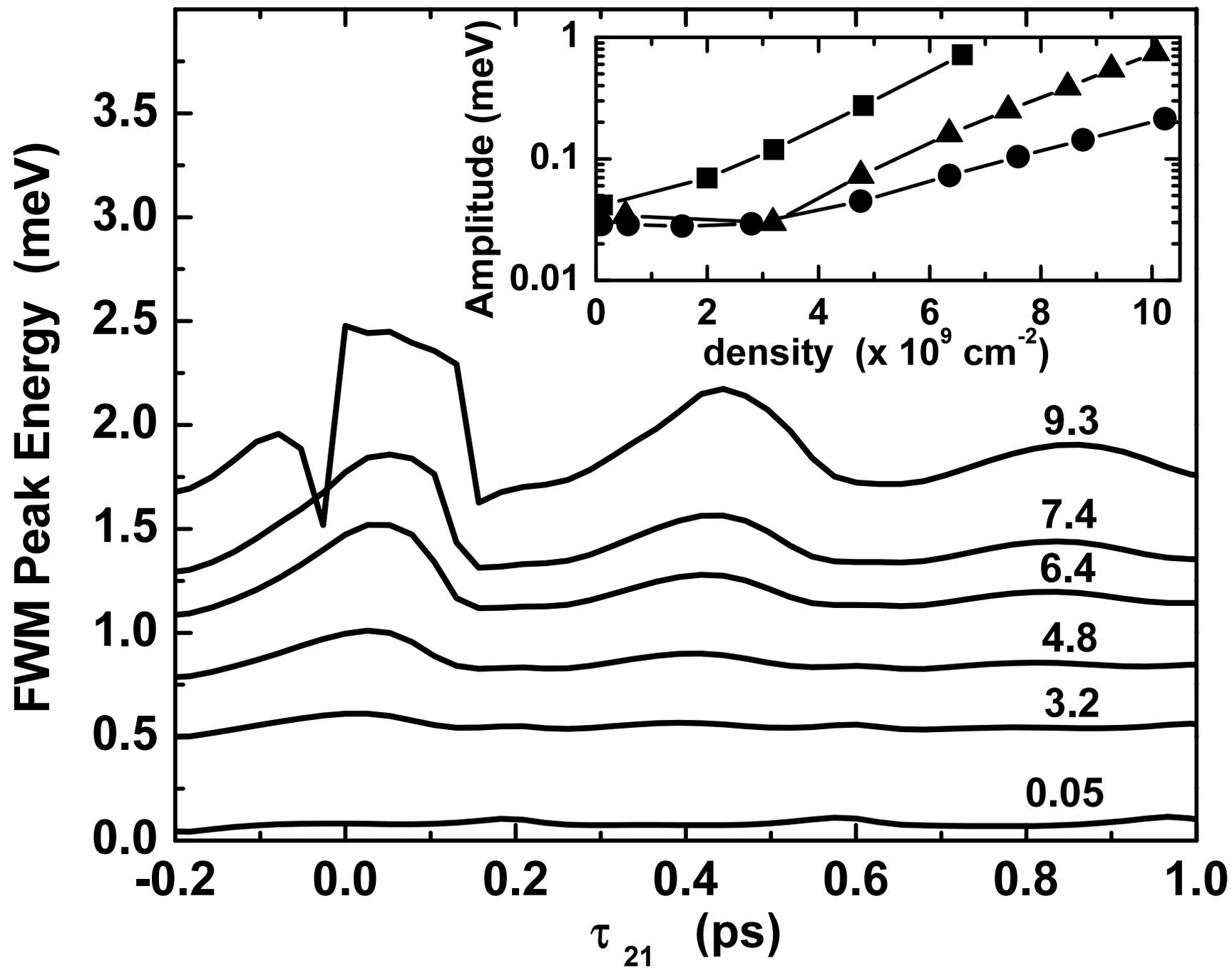
\item FIG. 3.\qquad The $\mu =-1$ SRFWM peak energy relative to $\hbar
\omega _{-1}$ versus $\tau _{21}$ at $\omega _{c}=\omega _{0}-2.27\omega _{B}
$ for a series of densities. \ The curve offsets are a real effect due
intraband renormalization of the DC field. \ The inset shows peak
oscillation amplitude as a function of density for $\left( \omega
_{c}-\omega _{0}\right) /\omega _{B}=-1.24$ (squares), $-2.27$ (circles) and 
$-2.83$ (triangles).
\end{description}


\begin{thebibliography}{99}
\bibitem{PSFcomment} We only consider areal exciton densities $\leq $ $%
10^{10}/cm^{2}$. \ Simple arguments and explicit calculations \cite%
{DignamHawton03} show that PSF for 1s excitons is unimportant if the
densities is much less than the inverse of the exciton area. \ For GaAs,
this density is $2\times 10^{11}/cm^{2}$, and hence PSF will certainly be
negligible for the densities of interest.
\bibitem{Tcomment} We find that finite population decay times have little
qualitative effect on the results presented here.
\end{thebibliography}
\end{document}